\documentclass{elsart}
\usepackage{graphicx}
\usepackage{amsmath}
\usepackage{amssymb}
\usepackage[german,english]{babel}
\begin{document}
\sloppy
\begin{frontmatter}
\frenchspacing
\title{Stability Analysis and Stabilization Strategies for Linear Supply Chains}

\author[a]{Takashi Nagatani}
\author[b]{Dirk Helbing}
\address[a]{Department of Mechanical Engineering,
Shizuoka University, Hamamatsu 432-8561, Japan}
\address[b]{Institute for Economics and Traffic, 
Dresden University of Technology, 01062 Dresden, Germany}

\begin{abstract}
Due to delays in the adaptation of production or delivery rates, supply chains
can be dynamically unstable with respect to perturbations in the consumption
rate, which is known as ``bull-whip effect''. Here, we study several conceivable production strategies
to stabilize supply chains, which is expressed by different specifications of the management
function controlling the production speed in dependence of the stock levels. In particular, we
will investigate, whether the reaction to stock levels of other 
producers or suppliers has a stabilizing effect. We will also demonstrate that the anticipation of
future stock levels can stabilize the supply system, given the forecast horizon $\tau$ is long enough.
To show this, we derive linear stability conditions and carry out simulations for different
control strategies. The results indicate that the linear stability analysis is a helpful tool for
the judgement of the stabilization effect, although unexpected deviations can occur in the
non-linear regime. There are also signs of phase transitions and chaotic behavior, but this
remains to be investigated more thoroughly in the future.
\end{abstract}
\begin{keyword}
Supply chains, bull-whip effect, stop-and-go traffic, stability analysis, non-linear dynamics, 
phase transitions, stabilization strategies, forecasts\\[3mm]
PACS: 
89.65.Gh, 
47.54.+r, 
89.40.+k, 
89.20.Bb 
\end{keyword}
\end{frontmatter}

\section{Introduction}

Concepts from statistical physics and non-linear dynamics have been very successful in
discovering and explaining dynamical phenomena in traffic flows \cite{reviews}. Many of these phenomena
are based on mechanisms such as delayed adaptation to changing conditions and 
competition for limited resources, which are relevant for production systems as well. 
Therefore, mathematicians
\cite{Armbruster}, physicists \cite{Hel}, traffic scientists \cite{Daganzo}, and economists \cite{Witt} 
have recently pointed out that methods used for the investigation of traffic dynamics
are also of potential use for the study of supply networks.
\par
In the following, we will investigate the stability of a linear supply chain. 
The assumed model consists of a series of $u$ suppliers $i$, which 
receive products from the next ``upstream'' supplier $i-1$ and generate products
for the next ``downstream'' supplier $i+1$ (see Fig.~1). The final products are delivered to the
consumers $u+1$. Here, we will assume that the consumption rate is subject to perturbations,
which may cause variations in the stock levels and deliveries of upstream suppliers. This is due to delays in
their adaptation of the delivery rate. Under certain conditions, the oscillations in delivery 
and in the resulting inventories (stock levels of the products) grow from one 
producer to the next upstream one. This is called the bullwhip effect \cite{Daganzo} and reported,
for example, for beer distribution \cite{beer1,beer2}. Similar dynamical effects
are also known for a series of subsequent production processes.
\par
In the following, we will investigate this amplification effect for a simple supply chain model 
by means of linear stability analysis and computer simulations. Interestingly, special cases of this model
turns out to have common features with some traffic models \cite{Hilliges,Bando}. 
However, production systems have more general
features, which are still to be explored. For example, 
\begin{itemize}
\item one may have complex supply networks rather than one-dimensional spatial interactions, and 
\item the adaptation strategies of the producers can be varied
to a much larger extent than those of drivers. 
\end{itemize}
While the first aspect is discussed in Refs.~\cite{Hel,Seba}, in this contribution  
we will focus on the latter aspect, in particular
the possibility to stabilize production by taking into account the inventories of other suppliers
and by anticipating changes in the inventories. In Sec.~\ref{Mod}, we will formulate the
supply chain model, while in
Sec.~\ref{Lin}, we will derive conditions for the linear stability of it. 
Section~\ref{Sim} introduces our simulation approach, while Sec.~\ref{Strat} will
compare different management strategies (``policies'') with regard to their effects on the
time-dependent product flows. Section~\ref{Conc} finally summarizes our findings,
discusses them in a broader context, and indicates further research directions.

\section{The Supply Chain Model} \label{Mod}

We will assume $u$ suppliers $i$ delivering their product(s) to the next downstream supplier $i+1$
\cite{prep}. The stock level (``inventory'') at supplier $i$ shall be denoted as $N_i$. It changes in time $t$
according to the equation
\begin{equation}
\label{eq3}
\frac{dN_i }{dt} = \lambda_i (t) - \lambda_{i + 1} (t) 
\end{equation}
\cite{Lefeber}. Here, $\lambda_i$ has the meaning of the rate at which supplier $i$ receives 
ordered products from supplier $i-1$, while $\lambda_{i+1}$ is the rate at which he delivers
products to the next downstream supplier $i+1$. Therefore,
Eq.~(\ref{eq3}) is just a continuity equation which reflects the conservation of the 
quantity of products. (It is easy to generalize this equation to cases where products are lost.
One would just have to add a term of the form $-\gamma_i N_i(t)$. It is also possible to
treat cases, where products are broken down into smaller units or several units are
required to produce one unit of a new product. This is discussed in Ref.~\cite{Hel}.)
Boundary conditions must be formulated for $i=0$, which corresponds to the supplier of
the raw materials (fundamental resources), and for $i=u+1$, which corresponds to the consumers. 
That is, $\lambda_0$ is the supply or production rate of the basic product, while $\lambda_{u+1}$
is the consumption rate. These are specified in Sec.~\ref{Sim}. 
\par
The question remains, how the delivery rates $\lambda_i$ evolve in time. It is reasonable
to assume that the temporal change of the delivery rate is 
proportional to the deviation of the actual delivery rate from the desired one $W_i$
(the order rate) and its adaptation takes on average some time interval $T$. According to this, we have
the equation
\begin{equation}
\label{eq4}
\frac{d\lambda _i }{dt} = \frac{1}{T}\left[ {W_i(t)  - \lambda _i (t)} 
\right] \, .
\end{equation}
The order rate $W_i$ will usually be reduced with increasing stock levels $N_j$,
but their temporal changes $dN_j/dt$ may be taken into account as well, e.g. when
the stock levels are forecasted. Therefore, it is natural to assume a general dependence of the form
\begin{equation} 
W_i(t) = W_i\left( {\left\{ {N_j (t)} 
\right\},\left\{ {dN_j (t) / dt} \right\}} \right) \, . 
\end{equation}
The function $W_i$ reflects the management strategy, i.e. the
order policy regarding the desired delivery rate as a function of the actual stock levels
$N_j(t)$ or anticipated stock levels $N_j(t) + \tau dN_j(t)/dt\approx N_j(t+\tau)$ 
(in first order Taylor approximation). 
The simplest strategy of supplier $i$ would be to react to the own stock level $N_i(t)$. However, it may
be useful to consider also the stock levels of the next downstream suppliers $j> i$, as these
determine the future demand, and the stock levels of the next upstream suppliers $j< i$, as
they determine future deliveries or shortages of the product (``out of stock'' situations).
We will, therefore, assume 
\begin{equation}
W_i(t) = W_i\left( {\left\{ {N_j (t)} 
\right\},\left\{ {dN_j (t) / dt} \right\}} \right) = W(N_{(i)}(t)) \, ,
\end{equation}
where
\begin{equation}
 N_{(i)}(t) = \sum_{l=-n}^n c_l \left(N_{i + l} + \tau \frac{dN_{i + l}}{dt}\right)
\label{en}
\end{equation}
is a weighted mean value of the own stock level and the ones of the next $n$ upstream and $n$ downstream 
suppliers (with $2n+1 \le u$). For $i+l < 0$ and $i+l > u$, the weights $c_l$ are always set to zero, and they
are normalized to one:
\begin{equation}
\sum_{l=-n}^n c_l = 1 \, . 
\end{equation}
For $\tau = 0$, the management adapts the delivery rate $\lambda_i$ to the {\em actual} 
weighted stock level $N_{(i)}(t)$, while for $\tau > 0$, the management orients
at the {\em anticipated} weighted stock level. The parameter $\tau$ has the meaning of
the forecast time horizon, and the specification of the parameters
$c_k$ and $\tau$ reflects the management strategy. 
\par
It is reasonable to assume that some strategies will be more favourable than others. For example, some
strategies will produce oscillating supply patterns (the ``bull-whip effect''), while others will stabilize
the product flow. These aspects will be discussed in Sec.~\ref{Strat}. Here, we will first discuss some potential
applications: We have already mentioned beer distribution, for which the bull-whip effect
has been impressively demonstrated. Similar effects are expected for all delivery chains, which
contain enough levels of delivery. We actually do not have to assume a linear structure, as
a hierarchical distribution system (with, for example, one producer, a few huge global suppliers,
some big country-based suppliers, more medium-sized regional suppliers, and many local suppliers)
will have similar properties, if the suppliers belonging to the same level behave approximately
the same (see Ref.~\cite{Hel}, Figs. 10 and 11). Linear supply systems are, however, known from 
transport chains. These are, for example, used in disaster management (and often made up 
of several dozen or hundred persons). Moreover, some production processes have similar features
as well, in particular if they use bucket brigades \cite{Arm}. 
In this case, the index $i$ represents the different successive production steps or machines,
$\lambda_i$ describes the corresponding production rate, and the so-called 
production function $W_i$ reflects the desired production rate as a function of the
stock levels $N_i$ in the respective output buffers. Although the above model does not consider
effects of limited buffer sizes (full buffers) \cite{Peters} or of limited transport capacities, these
features can be easily added (see Ref.~\cite{Hel}).
\par
Before we continue with a linear stability analysis of the above model, 
let us shortly point out some analogies with other systems. For the particular
specifications $T\rightarrow 0$,
$N_{(i)}(t) = N_i(t)$, and $W_i(t) = N_{i-1}(t) V(N_i(t))$, the mathematical structure of the
supply chain model agrees with the macroscopic traffic model of Hilliges and Weidlich \cite{Hilliges}:
\begin{equation}
 \frac{d N_i(t)}{dt} = N_{i-1}(t) V(N_i(t)) - N_i(t) V(N_{i+1}(t)) \, .
\end{equation}
However, in this equation, $N_{i-1}$ has the meaning of the traffic density in a road section $i-1$ 
of length $\Delta x$, while $V(N_i)$ is the average velocity-density relation of vehicles on
road section $i$. In this model, vehicles in section $i-1$ are assumed to adapt to the velocity in the next
road section $i$, which describes an anticaptive driving behavior.
\par
Another special case is accidentally related to the so-called optimal velocity model \cite{Bando}. This
car-following model assumes an acceleration equation of the form
\begin{equation}
 \frac{dv_i(t)}{dt} = \frac{V_{\rm o}(d_i(t)) - v_i(t)}{T} 
\end{equation}
and the complementary equation
\begin{equation}
 \frac{dd_i(t)}{dt} = - [v_{i}(t) - v_{i+1}(t)] \, .
\end{equation}
In this ``microscopic'' traffic model, the quantities have a totally different meaning
than in the above supply chain model: The index
$i$ represents single vehicles, $v_i(t)$ is their actual velocity of motion, $V_{\rm o}$ the so-called
optimal (safe) velocity, which depends on the distance $d_i(t)$ to the next vehicle ahead, and
$T$ is an adaptation time. Comparing this equation with Eq.~(\ref{eq4}), the velocities $v_i$
would correspond to the delivery rates $\lambda_i$, the optimal velocity $V_{\rm o}$ to the
order rate (desired delivery rate) $W_i$, and the inverse vehicle distance $1/d_i$ would approximately
correspond to the stock level $N_i$ (apart from a proportionality factor). 
This shows, that the analogy between supply chain and traffic
models concerns only their mathematical structure, but not their interpretation, although both
relate to transport processes. Nevertheless, this mathematical relationship can give us hints,
how methods, which have been successfully applied to the investigation of traffic models before,
can be generalized for the study of supply networks. While in traffic flow,
the velocity-density relation is empirically given, the new feature of supply chains is that the management
has a large degree of freedom to choose the order function $W_i$. In principle, this can be specified
not only as a function of the own stock level, but of the stock levels of other suppliers as well, if the information is available.
Therefore, instead of the simple dependence of $W_i$ on $N_i$, we have to explore the considerably
more complex dependence on the weighted stock level (\ref{en}). This will lead to new results
regarding the dynamical behavior, which are of potential relevance for the management
of supply chains. 

\section{Linear stability analysis} \label{Lin}

Let us now study the stability of the steady-state solution of Eqs. (\ref{eq3}) and (\ref{eq4}). 
This steady-state solution is given by $N_i = N_0$ and $\lambda_i = \lambda _0 = W(N_0)$. 
By $\delta N_i (t)$ and $\delta \lambda _i (t)$,
we will denote small deviations from the steady state: 

\begin{equation}
\label{eq6}
N_i (t) = N_0 + \delta N_i (t),
\end{equation}

\begin{equation}
\label{eq7}
\lambda _i (t) = \lambda _0 + \delta \lambda _i (t).
\end{equation}

Then, the linearized equations obtained from Eqs. (\ref{eq3}) and (\ref{eq4}) read
\begin{equation}
\label{eq8}
\frac{d\delta N_i (t)}{dt} = \delta \lambda _i (t) - \delta \lambda _{i + 1} 
(t)
\end{equation}
and
\begin{equation}
 T\frac{d\delta \lambda _i (t)}{dt} = W'(N_0 ) 
\sum_{l=-n}^n c_l \left(\delta N_{i + l} + \tau \frac{d\delta N_{i + l}}{dt} \right) 
- \delta \lambda _i (t) , 
\label{eq9}
\end{equation}
where $W'(N_0 )$ is the derivative of the control function at $N_{(i)} = N_0$.

With the ansatz $\delta N_j (t) = X\exp ({\rm i}jk + zt)$ and $\delta \lambda _j (t) 
= Y\exp ({\rm i}jk + zt)$, where ${\rm i} = \sqrt{-1}$ denotes the imaginary unit,
$k$ the wave number, and $z$ a complex eigenvalue, one obtains
\begin{equation}
\label{eq10}
zX = (1 - e^{{\rm i}k})Y
\end{equation}
and
\begin{equation}
 zTY = W'(N_0 ) (1 + z\tau )
\left( \sum_{l=-n}^n c_l e^{{\rm i}kl}
\right) X - Y . 
\label{eq11}
\end{equation}

Inserting Eq. (\ref{eq10}) into Eq. (\ref{eq11}), we find

\begin{equation}
 z^2T = W'(N_0 )(1 + z\tau )(1 - e^{{\rm i}k}) \sum_{l=-n}^n c_l e^{{\rm i}kl}
- z. 
\label{eq12}
\end{equation}
This quadratic equation in $z$ may be solved with respect to $z$. In order to investigate the limiting case of
small wave numbers $k$ (large wave lengths), we will use the expansion of the exponential function:
\begin{equation}
 e^{{\rm i}kl} = \sum_{j=0}^\infty \frac{({\rm i}kl)^j}{j!} 
 = \sum_{j=0}^\infty \frac{l^j}{j!} ({\rm i}k)^j \, .
\end{equation}
Inserting this into Eq. (\ref{eq12}), one finds that the leading term of $z$ is of the order of 
${\rm i}k$, when ${\rm i}k \to 0$, $z \to 0$. Let us derive the long wave expansion of $z$, 
which is determined order by order around ${\rm i}k \approx 0$. By expanding 
\begin{equation}
z =  \sum_{j=1}^\infty z_j ({\rm i}k)^j \, ,  
\end{equation}
the first- and second-order terms of ${\rm i}k$ are obtained as
\begin{equation}
\label{eq13}
z_1 = - W'(N_0 )
\end{equation}
and
\begin{eqnarray}
 z_2 &=& - z_1{}^2T 
- W'(N_0 ) \left( z_1\tau + \frac{1}{2} + \sum_{l=-n}^n l c_l \right) \, .
\label{eq14}
\end{eqnarray}
If $z_2 $ is a negative value, the steady state becomes unstable for long 
wavelength modes, i.e. modes with small wave numbers $k$. 
However, if $z_2 $ is a positive value, the steady state is 
stable. Therefore, the linear stability condition is given by 
\begin{equation}
T  < \tau +
  \frac{1}{|W'(N_0 )|}\left( \frac{1}{2} + \sum_{l=-n}^n l c_l \right) \, ,
\label{eq15}
\end{equation}
where we have used $-W'(N_0) = |W'(N_0)| > 0$ \cite{prep}, 
as the order rate normally decreases with increasing stock levels $N_0$.
According to Eq.~(\ref{eq15}), anticipation of the stock levels ($\tau > 0$) has always a positive effect on stability,
and it appears to be favorable to orient the order rate at downstream suppliers (with $l>0$). These points will be
discussed in Sec.~\ref{Strat} in more detail. 
However, the validity of condition (\ref{eq15}) is restricted to systems with a large number $u \gg n$ of 
suppliers. For small systems, we may have corrections
due to boundary effects (see Ref.~\cite{finite} for their treatment). Generally, smaller systems tend to be
more stable.

\section{Simulation Approach} \label{Sim}

Apart from a linear stability analysis, we have carried out computer simulations 
of Eqs. (\ref{eq3}) and (\ref{eq4}), using Euler integration with a time discretization of
$\Delta t = 0.01$. In order to make boundary effects small, we have chosen a
large number $u=200$ of suppliers, which is realistic for the transport chains
mentioned in Sec.~\ref{Mod}. Moreover, the boundary conditions have been specified as follows:
\begin{equation}
\lambda _0 = W(N_0) \qquad \mbox{and} \qquad N_{u + 1} (t) = N_0 + \xi (t) / 2 \, ,
\end{equation}
where $\xi (t)$ is a white noise with mean value
\begin{equation}
 \langle \xi (t) \rangle = 0
\end{equation}
and time correlation 
\begin{equation}
 \langle \xi(t)\xi (t') \rangle = \delta _{tt'} / 4 .
\label{eq24}
\end{equation}
Finally, the order rate (i.e. the desired supply rate) was chosen as
\begin{equation}
 W(N_{(i)}) = 1 - [ \tanh(N_{(i)} - N_{\rm c}) + \tanh(N_{\rm c})]/2 \, .
\label{eq5}
\end{equation}
This function can be viewed as representative
for many other monotonously falling functions with a turning point $N_{\rm c}$. Here,
the turning point was set to $N_{\rm c} = 3$, which guarantees $N_i(t) \ge 0$ for small enough
adaptation times $T$. The fact that the values of $W$ 
range from 1 to 0 is no restriction: Every bounded and monotonously falling
function with a minimum value of 0 can be scaled so that it varies from
1 to 0. This just implies a scaling
of the delivery rates $\lambda_i$, i.e. a time scaling. In addition, for many order functions $W$,
the quantities $N_i$ of products can be scaled in a way that they are 
approximated by relation (\ref{eq5}). Although other kinds of functions are
conceivable as well (see e.g. Ref.~\cite{Hel}), we expect quantitatively the same findings
in the linear regime around the stationary state and qualitatively the
same phenomena in the nonlinear regime (for many but not all monotonously falling functions).

\section{Discussion of Different Control Strategies}\label{Strat}

In the following, we will investigate some reasonable management strategies (order strategies) in more detail.
In particular, we will study how the dynamics of a linear supply chain can be stabilized
by selecting an appropriate strategy. We will set $c_{-l} =0$ for all $l> 0$, as 
positive values would reduce the stability threshold according to Eq.~(\ref{eq15}). 
This means, it is not helpful to take into account the stock levels of upstream
suppliers $j = {i-l}$ in the order strategy $W$, as it is expected to destabilize the supply
chain. Consequently, we will only discuss order strategies considering the own stock
level and the ones of downstream suppliers:

\begin{itemize}
\item[(A)] With strategy A we mean the case that the order policy of the management of supplier
$i$ takes into account 
only the own present stock level $N_i(t)$. That is, strategy A does not consider anticipation ($\tau = 0$),
and the weigthed stock level agrees with the own one: $N_{(i)}(t) = N_i(t)$. This in inserted
into Eq.~(\ref{eq5}).

The linear stability condition for this order strategy reduces to 
\begin{equation}
 |W'(N_0 )| <  \frac{1}{2T} \, ,
\label{Eq17}
\end{equation}
which is mathematically equivalent with the stability condition of the optimal velocity model
mentioned before \cite{Bando}. Correspondingly, if the change $W'(N_0)$ of the order rate
with the stock level $N_i$ at the stationary point $N_i = N_0$ is greater than 
half of the inverse adaptation time $1/T$, the supply chain is expected to behave unstable.
In this situation, small fluctuations in the consumption rate can trigger large oscillations
in the stock levels. 

\item[(B)] In order to suppress these oscillations, with strategy B, which corresponds to
the specification $N_{(i)}(t) = N_i (t) + \tau dN_i (t) / dt$,
we have additionally taken into account the effect of anticipation of the stock levels.
That is, with strategy B, the management reacts to the forecasted stock level at time
$t + \tau $, which presupposes that the (sufficiently reliable) forecast time horizon is at least $\tau$.
The corresponding stability condition becomes
\begin{equation}
\label{eq16}
 T < \tau + \frac{1}{2|W'(N_0 )|} \, .
\end{equation}
According to this, the neutral stability line is shifted to higher values by the time horizon $\tau$ of the forecast,
and long enough forecast time horizons $\tau$ should always stabilize the supply
chain. Due to Eq.~(\ref{eq15}), this is also true in combination with other strategies than A.
However, in reality the stabilization strength is, of course, limited by the practical forecast capability.
Nevertheless, anticipation should always have a stabilizing effect.
\par
Figure~2 shows plots of the 
neutral stability lines against the steady-state inventories $N_0$ for $\tau = 
0.0$, 0.4, 0.8, 1.2, and 1.6, where $N_{\rm c} = 3$. 
Note that the supply chain behaves stable below the neutral 
stability line, while it behaves unstable above it. With increasing 
$\tau $, the neutral stability line increases.

In Figure~3, we have compared strategies A and B with respect to the
time evolution of the stock levels (inventories) $N_i$ of all suppliers $i$. 
While subfigure (a) corresponds to the dynamical pattern of strategy A with adaptation time 
$T=2$ and anticipation time horizon $\tau = 0$, the dynamical pattern of 
strategy B for $T=2$ and $\tau = 0.4$ is shown in (b). Without forecasting (see Fig.~3a), the
stock levels of most suppliers show significant oscillation amplitudes, 
while with anticipation, the oscillations are weaker and 
concern mainly the suppliers upstream of $i\approx 150$.
Note that the oscillations propagate backwards at a constant speed, which is analogous to the
propagation of stop-and-go waves \cite{stopgo}.

For strategy B, let us now study 
the dependence of the amplitude of the inventories $N_i$ on the anticipation time horizon 
$\tau$. Figure~4 shows the plot of the maximum oscillation amplitude as a function of the 
time horizon $\tau $ for an adaptation time $T=2$. With 
increasing time horizon $\tau$, the amplitude decreases. At a critical threshold of about $\tau = 
0.9$, the oscillations disappear completely, although the anticipation time horizon is considerably 
smaller than the adaptation time $T$ (which is due to the finite value of
$|W'(N_0 )|$). For higher values than the critical threshold $\tau = 0.9$, which can be estimated by
the linear stability condition, the supply chain does not oscillate and is stable. 

\item[(C)] When the practical forecast time horizon is too small to stabilize the supply chain,
anticipation should be combined with better strategies than A. It is reasonable that the
consideration of the stock level of the next downstream supplier can improve the stability,
if the information is somehow available. Therefore, with strategy C we assume that the management 
also takes into account the actual inventory of the downstream (front) nearest-neighbor 
$i+1$ it delivers to. In mathematical terms, we set $N_{(i)}(t) = c_0 N_i (t) + (1 - c_0 )N_{i + 1} (t)$
with a model parameter $c_0$ ($0 \le c_0 \le 1$). 
Instead of $ T <  0.5 / |W'(N_0 )|$ (see Eq.~(\ref{Eq17})), the related linear stability condition becomes
\begin{equation}
\label{eq17}
T < \frac{1 + 2(1 - c_0 )}{2|W'(N_0 )| } \, . 
\end{equation}
Therefore, the stability threshold is increased for $c_0 < 1$. The case $c_0 = 1$ corresponds exactly
to strategy A (our reference strategy for the purpose of comparison).

Figure 5 shows plots of the neutral stability point against the steady-state 
inventory for $c_0 = 1.0$, 0.8, 0.6, 0.4, and 0.2, where $N_{\rm c} = 3$. Each solid 
curve represents a neutral stability line. Again, the production system is stable 
below the neutral stability line, while the system is unstable above the curve. 
With decreasing $c_0 $, the neutral stability line increases. As a consequence, 
the production system is stabilized by taking into account the inventory of 
the next downstream supplier. 

We have also studied the dependence of the oscillation amplitude on the adaptation 
time for strategy C with $c_0 = 0.7$ (see Fig.~6). The oscillation of the inventories appears at
about $T=1.8$ and strongly grows with increasing adaptation time $T$. 

\item[(D)] As the consideration of the stock level of the next downstream supplier has 
a stabilization effect, we will investigate with strategies D and E, whether it is even
better to take into account the stock levels $N_j(t)$ of the second-next supplier ($j=2$) or the consumer market
($j=u+1$), given this information
is available. For strategy D we assume that the management considers the inventory of the 
second-next (instead of the next) downstream supplier: $N_{(i)}(t) = c_0 N_i (t) + (1 - c_0 )N_{i + 2} (t)$. 
In this case, one obtains the linear stability condition
\begin{equation}
\label{eq18}
T < - \frac{1 + 4(1 - c_0 )}{2W'(N_0 )} \, ,
\end{equation}
which implies an increased stability threshold for $c_0 < 1$, i.e. a greater stability.

Figure 7 shows plots of the neutral stability lines against the steady-state 
inventory for $c_0 = 1.0$, 0.8, 0.6, 0.4, and 0.2, where $N_c = 3$. The value
$c_0 = 1$ again corresponds to the reference strategy A.
With decreasing $c_0 $, the neutral stability line increases. For the same 
value of $c_0 $, strategy D is more stable than strategy C. As a 
result, the production system is more efficiently stabilized by taking into 
account the inventory of the second-next downstream supplier.

\item[(E)] Moreover, with strategy E, the management is assumed to take into account the consumption 
directly, i.e. $N_{(i)}(t) = c_0 N_i(t) + (1 - c_0 )N_{u + 1}(t) $ in control function (\ref{eq5}),
where $N_{u + 1}(t)$ represents the boundary condition. One would assume that this
strategy would have a greater stabilization effect than strategies C and D. In order to judge
this, let us compare for strategies C, D, and E the profiles
of the inventories of all suppliers  at time $t=20000$, which are displayed
in Figure 8. The chosen parameter values are $T=2$ 
and $N_c = 3$. While plot (a) shows the inventories for
strategy C with $c_0 = 0.9$, plot (b) displays the inventories 
for strategy D with the same value of $c_0$.  Moreover,
plot (c) shows the inventories for $c_0 = 0.8$ and strategy D', for which we assume for comparison 
that a manager considers the stock-level of the third-next downstream supplier:
$N(t) = c_0 N_i (t) + (1 - c_0 )N_{i + 3} (t)$. Finally, plot (d) 
displays the inventories for strategy E with $c_{u + 1} = 1 - c_0 = 0.3$. Obviously, taking 
into account the inventories of downstream suppliers can reduce the oscillations, 
see (a) and (b). However, with strategy D', the supply chain may surprisingly
become more unstable, see (c). That is, if the management chooses 
an order strategy taking into account the inventory $N_{i + n}(t)$ of a further downstream supplier
$j=i+n$ with $n \ge 3$, the supply chain may be destabilized. This instability 
is in contrast to our expectations based on the linear stability analysis. The likely reason
are resonance effects in the non-linear regime (when the vicinity of the stationary state $N_i = N_0$ is left), 
so that the implications of the linear stability analysis have to be considered with care and checked by simulations.

Despite such kinds of surprises, strategy E exhibits a stabilization effect, since the 
oscillations are reduced, when the consumption is taken into account, see (d). 
However, the stabilization effect is not as strong as expected. To see this, let us investigate
the stabilization of strategies C, D, and E by means of Figure 9. It
shows the plots of the maximum oscillation amplitude of the 
inventories as a function of $1 - c_0$ for strategies C, D, and E, where $T=2$ and $N_c = 3$. With increasing 
fraction $1 - c_0 $, the maximum amplitude decreases for all three strategies. 
The oscillation of the inventories disappears at about $c_0 = 0.8$ for
strategy D. For strategies C and E, the oscillation disappears
at $c_0 = 0.6$ and 0.55, respectively. For strategy D, the reduction of the oscillation amplitude is
strongest. Surprisingly, 
the stabilization effect of strategy E is less than that of strategy C, which is probably due to
a resonance effect, similar to the observation for strategy D'.
\end{itemize}

\section{Summary, Discussion, and Outlook} \label{Conc}

{\rm Similar to stop-and-go waves in vehicle traffic,}
linear supply chains sometimes suffer from the bull-whip effect, i.e. growing oscillations
in the inventories (stock levels), which are due to delays in the adaptation of the delivery rate or production speed.
We have studied the impact of various reasonable management strategies with regard
to the stabilization or destabilization of the dynamics of a linear supply chain. 
This has been judged by the linear stability condition we derived, and 
the dynamical behavior of the inventories has additionally been 
investigated by means of simulations. While we have focussed on the effect of perturbations in
the consumption rate, the result of perturbations in the delivery rates of suppliers may be 
studied in a similar way. However, according to the linear stability analysis, we do not expect
any qualitatively new results.

We have shown that a supply chain can be stabilized, i.e. oscillations in the delivery rates and stock levels
can be reduced by anticipation of the temporal evolution 
of the inventories and by taking into account the inventories of other 
suppliers. In the case of perturbations
in the consumption rate, our results were as follows:
\begin{itemize}
\item Anticipation of the own future inventory was an efficient means
to stabilize the production system. Surprisingly, it turned out that anticipation time horizons considerably
smaller than the adaptation time were sufficient to reach complete stability. Similar results
are expected, if anticipated inventories of other suppliers are taken into account as
well. Fig.~4 suggests that the transition to stability with increasing anticipation time horizon
becomes discontinuous in the limit $u\rightarrow \infty$ of infinitely many
suppliers. 
It would be interesting to investigate this in more detail in the future, as well as the role of errors
in the forecast.
\item According to the linear stability analysis, the adaptation to a variation in the consumption rate
is better, if not only the own inventory,  but also the inventories of downstream 
suppliers or the consumer sector itself are taken into account by so-called {\em ``pull strategies''}. 
In contrast, considering the
inventories of upstream suppliers corresponding to {\em ``push strategies''}
destabilized the system (cf. Ref. \cite{FactPhys}). One would think that this is,
because the oscillations of upstream inventories tend to be larger, but the same result comes out from
a linear stability analysis, i.e. in the limit of vanishing oscillation amplitudes. It is rather the
direction of the information flow in the system which matters: The oscillations in the consumption rate
travel upstream, {\rm as in stop-and-go traffic \cite{stopgo}} (see Fig. 3).
\item The linear stability analysis gives a good idea under which conditions the oscillation amplitude
in the system becomes zero. For example, the results displayed in Figs.~4 and 9 are in good agreement
with our expectations. However, {\rm as for traffic systems,} the implications of a linear stability analysis 
are limited, because the evolving oscillation amplitudes are usually large, and non-linear effects 
dominate. For example, the emerging wave length in the system does often not agree with the most
unstable wave length mode, i.e. the mode with the largest growth rate \cite{Hab}. Therefore, the simulation
result displayed in Fig.~8(c) was not in agreement with our expectations based on the linear
stability analysis. Surprisingly, the oscillation amplitudes were larger, when inventories of further
downstream suppliers were taken into account. This point is related with the different
wavelength emerging in Fig.~8(c) compared to Fig.~8(a), (b), and (d), as the resulting oscillation amplitude
depends on the frequency \cite{Seba}. It would certainly be interesting to investigate in the future
the relation with period-doubling phenomena \cite{Peters}, which are known to exist for traffic systems \cite{double}.
\item Let us finally come back to the simulation results displayed in Fig.~9, showing the
amplitude of oscillations in the inventories for different management strategies. 
For $c_0 = 1$, strategies C, D, and E agree with strategy A, and the oscillation amplitudes are the same.
However, when in the weighted stock level $N_{(i)}(t) = c_0 N_i(t) + (1-c_0) N_j(t)$,
the weight $(1-c_0)$ of another inventory $N_j(t)$ is increased in the production strategy, there
is a surprise: The oscillation amplitudes are significantly reduced, when
the second next downstream supplier is taken into account with $N_j(t) = N_{i+2}(t)$ instead of
the next downstream one with $N_j(t) = N_{i+1}(t)$, as expected. However, considering the variation in the consumption
rate itself with $N_j(t) = N_{u+1}(t)$ has a very weak stabilization effect, although the consumer sector
is located even further downstream. This is probably due to resonance effects, when the
vicinity of the stationary state is left, and due to the fact that the characteristics of the perturbation
in the consumption rate differs from those of the emerging oscillation patterns.
\end{itemize}
In conclusion, there are several non-trivial and unexpected effects in the behavior of linear supply chains.
Therefore, simulation models describing the non-linear interactions and dynamics of supply
chains and production processes
could be relevant for their optimization. From the practical point of view it is, for example, useful
that Eq.~(\ref{eq15}) allows one to estimate the maximum adaptation time $T$ or the
minimum forecast time horizon $\tau$ supporting a stable supply chain. Moreover, the 
stabilizing effect of a reaction to inventories of downstream suppliers suggests to exchange these data 
on-line. Note that our conclusions regarding the stabilization by forecasts and the consideration
of downstream stock levels are expected to be transferable to more complex systems
than the linear supply chains treated here. They should be also applicable to cases where suppliers are 
characterized by different parameters, to situations with limited buffers and transportation capacities,
or to supply networks \cite{Seba}.

From the physical point of view, it will be particularly interesting
to study the conditions for period doubling phenomena in the future, and to investigate
whether the stabilization transition (when the weight $c_0$, the adaptation time $T$, or
the anticipation time horizon $\tau$ is varied) would become discontinuous in 
the limit $u\rightarrow \infty$ of infinitely many suppliers.
In view of the partially irregular patterns in Fig.~8, it will also be interesting to seek for conditions for
chaotic dynamics \cite{beer1,Chaos} and for concepts to control it \cite{control}.

\subsection*{Acknowledgments}
D.H. would like to thank Stefan L\"ammer for producing the schematic illustration displayed in Fig.~1.

\newpage
\section*{Figure captions}
FIG. 1. Illustration of the linear supply chain treated in this paper, including the key
variables of the model. Circles represent different suppliers $i$, $N_i$ their respective
stock levels, and $\lambda_i$ the delivery rate to supplier $i$ or the production speed
of this supplier. $i=0$ corresponds to the resource sector generating the basic products
and $i=u+1$ to the consumers.

\bigskip

FIG. 2. Plots of the neutral stability line as a function of the steady-state inventory
for  strategy B and various values of the anticipation time horizon (ATH)
$\tau = 0.0$, 0.4, 0.8, 1.2, and 1.6 ($N_c = 3$). The supply chain behaves stable 
below the neutral stability line, while it behaves unstable above the curve. Strategy A
corresponds to $\tau = 0$.

\bigskip

FIG. 3(a). Dynamics of the inventories of 200 suppliers $i$. (a) Strategy A 
with adaptation time $T=2$. (b) Strategy B with $T=2$ and anticipation time horizon
$\tau = 0.4$.

\bigskip

FIG. 4. Plot of the maximal amplitude of oscillation in the inventories as a function
of the time horizon $\tau $ for strategy B with adaptation time $T=2$.

\bigskip

FIG. 5. Plots of the neutral stability line as a function of the steady-state inventory for 
strategy C and $c_0 = 1.0$, 0.8, 0.6, 0.4, 0.2. Strategy A corresponds to $c_0 = 1$.

\bigskip

FIG. 6. Plot of the maximum amplitude of the inventories as a function of the adaptation time $T$ for
strategy C with $c_0 = 0.7$.

\bigskip

FIG. 7. Plots of the neutral stability line against the steady-state inventory for 
strategy D and $c_0 = 1.0$, 0.8, 0.6, 0.4, 0.2. Again, strategy A corresponds to $c_0 = 1$.

\bigskip

FIG. 8(a). Stock levels (inventories) of 200 suppliers $i$ for various strategies at time $t=20000$, where
the adaptation time is $T=2$ and the turning point is located at $N_c = 3$. 
(a) Inventories for strategy C with $c_0 = 0.9$. 
(b) Inventories for strategy D with $c_0 = 0.9$. (c) 
Inventories for strategy D' with $c_3 = 1 - c_0 = 0.2$. (d) 
Inventories for strategy E with $c_{u + 1} = 1 - c_0 = 0.3$.

\bigskip

FIG. 9. Plots of the maximum amplitude of oscillations in the inventories over $1 - c_0 $ for strategy C
(label $i+1$),  strategy D (label $i+2$), and strategy E (label $u+1$) with $T=2$ and $N_c = 3$.

\newpage
\begin{figure}[htbp]
 \begin{center}
  \includegraphics[width=12cm]{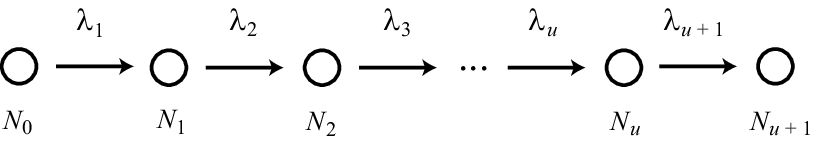}
 \end{center}
\caption[]{ }\label{Fig0}\end{figure}

\newpage

\begin{figure}[htbp]
 \begin{center}
  \includegraphics[width=10cm]{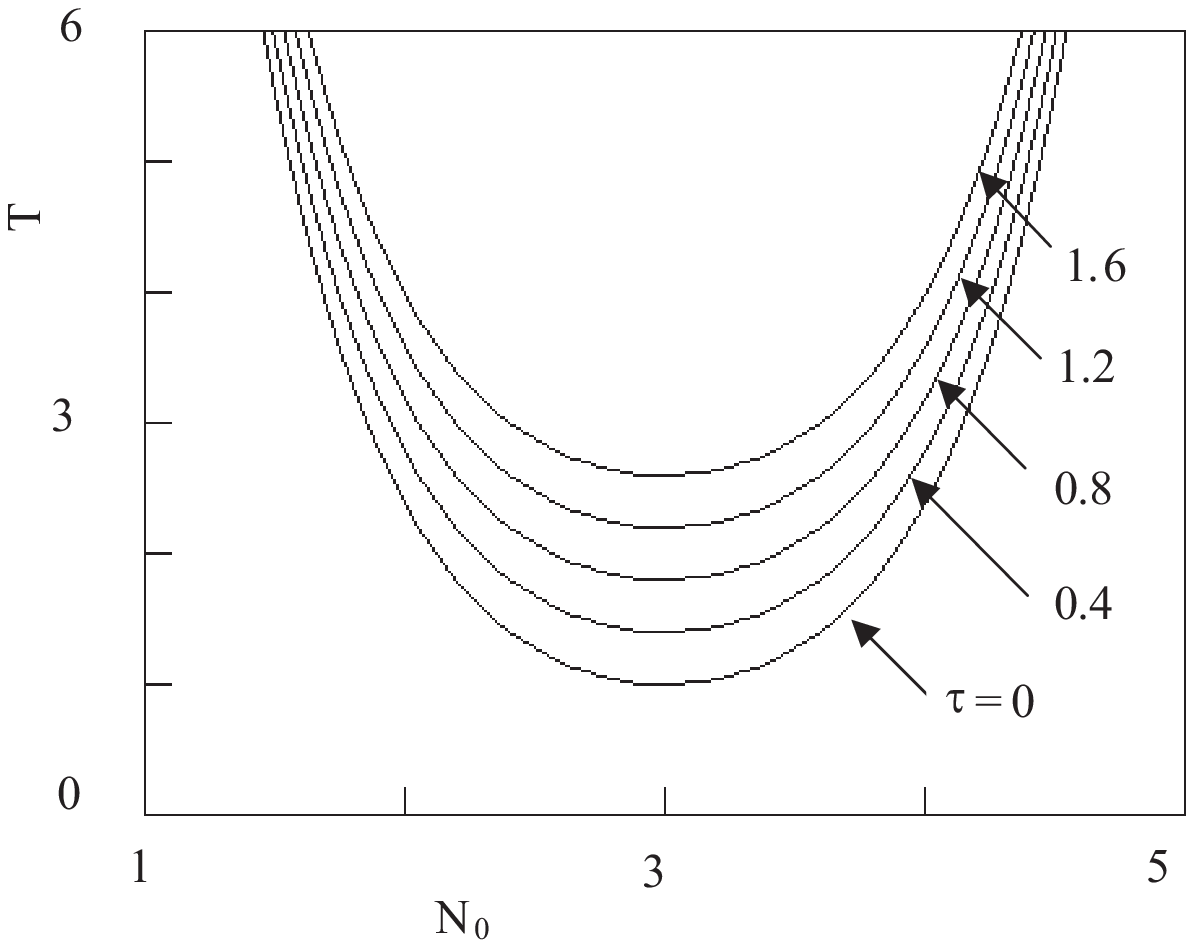}
 \end{center}
\caption[]{ }\label{Fig1}\end{figure}

\begin{figure}[htbp]
 \begin{center}
  \includegraphics[width=10cm]{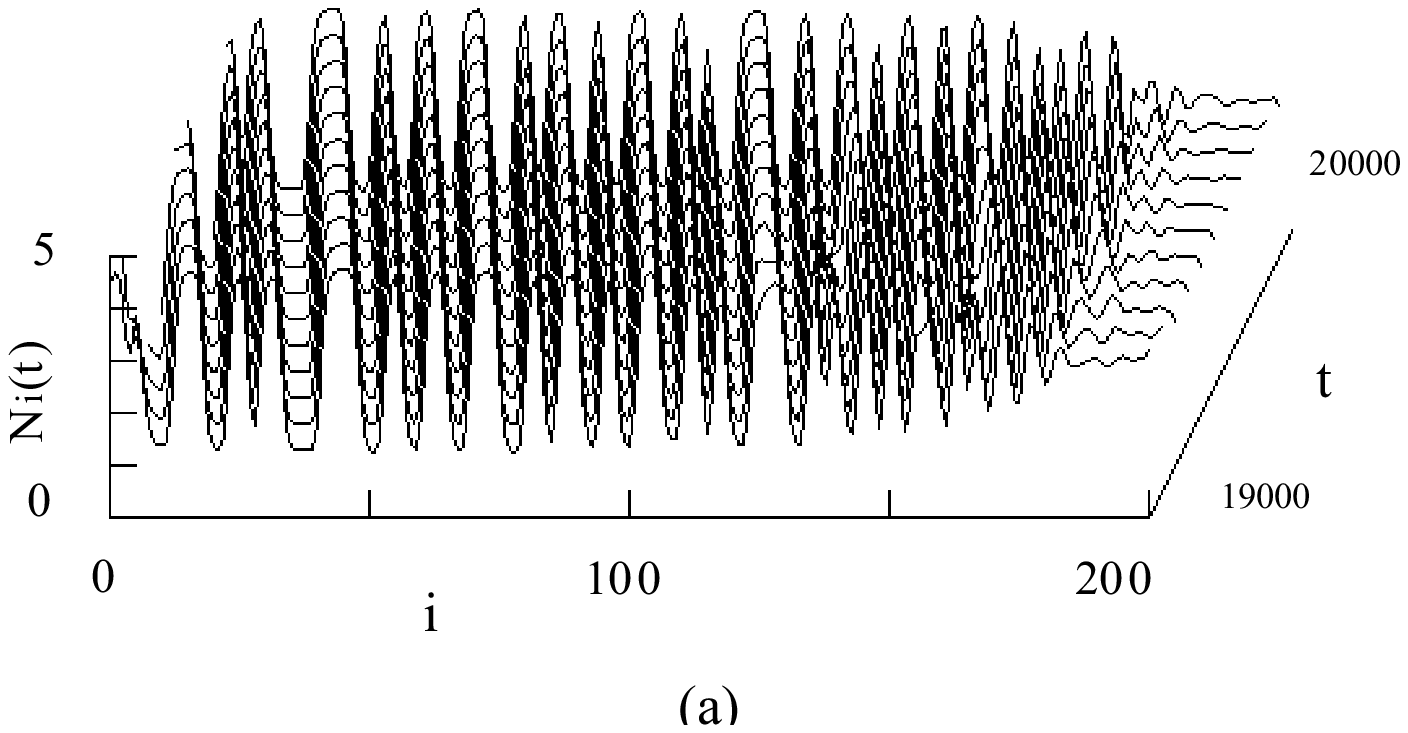}\\[-10cm]
%
  \includegraphics[width=10cm]{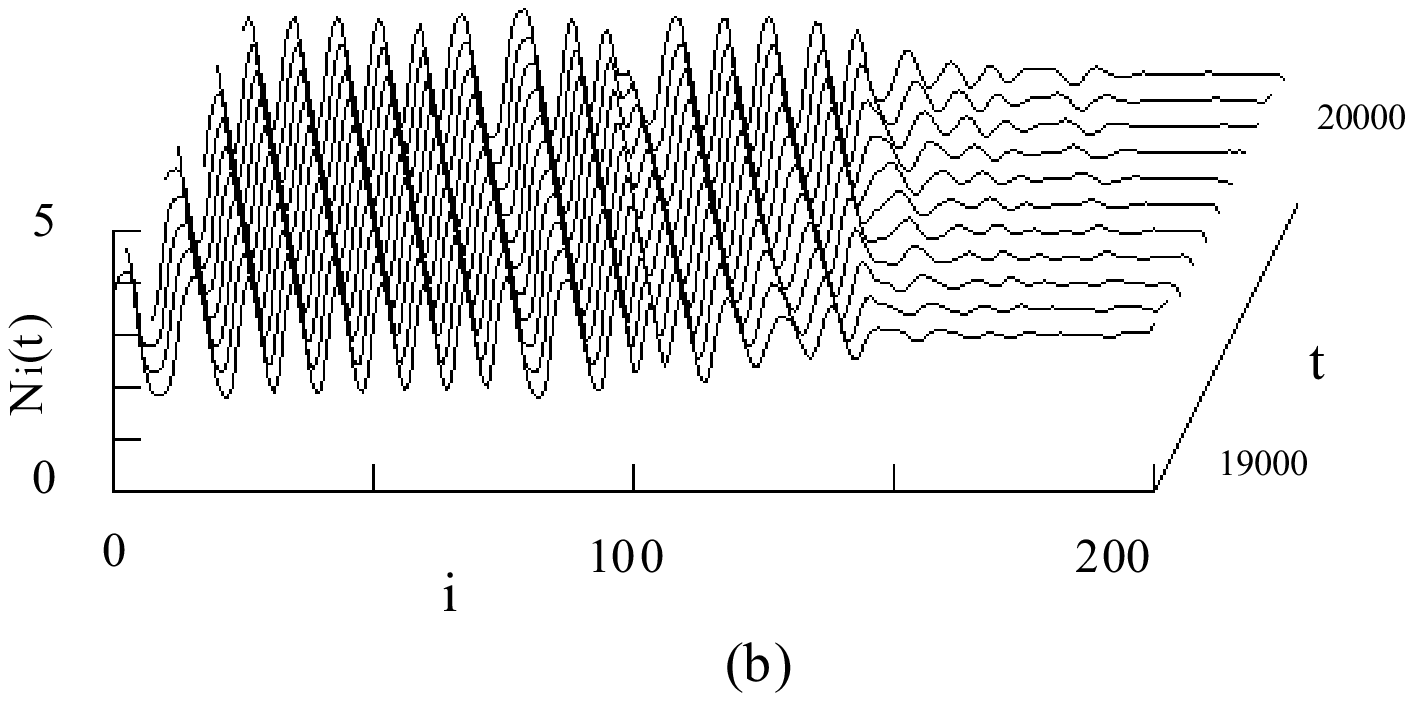}\\[-10cm]
 \end{center}
\caption[]{ }\label{Fig2}\end{figure}

\begin{figure}[htbp]
 \begin{center}
  \includegraphics[width=10cm]{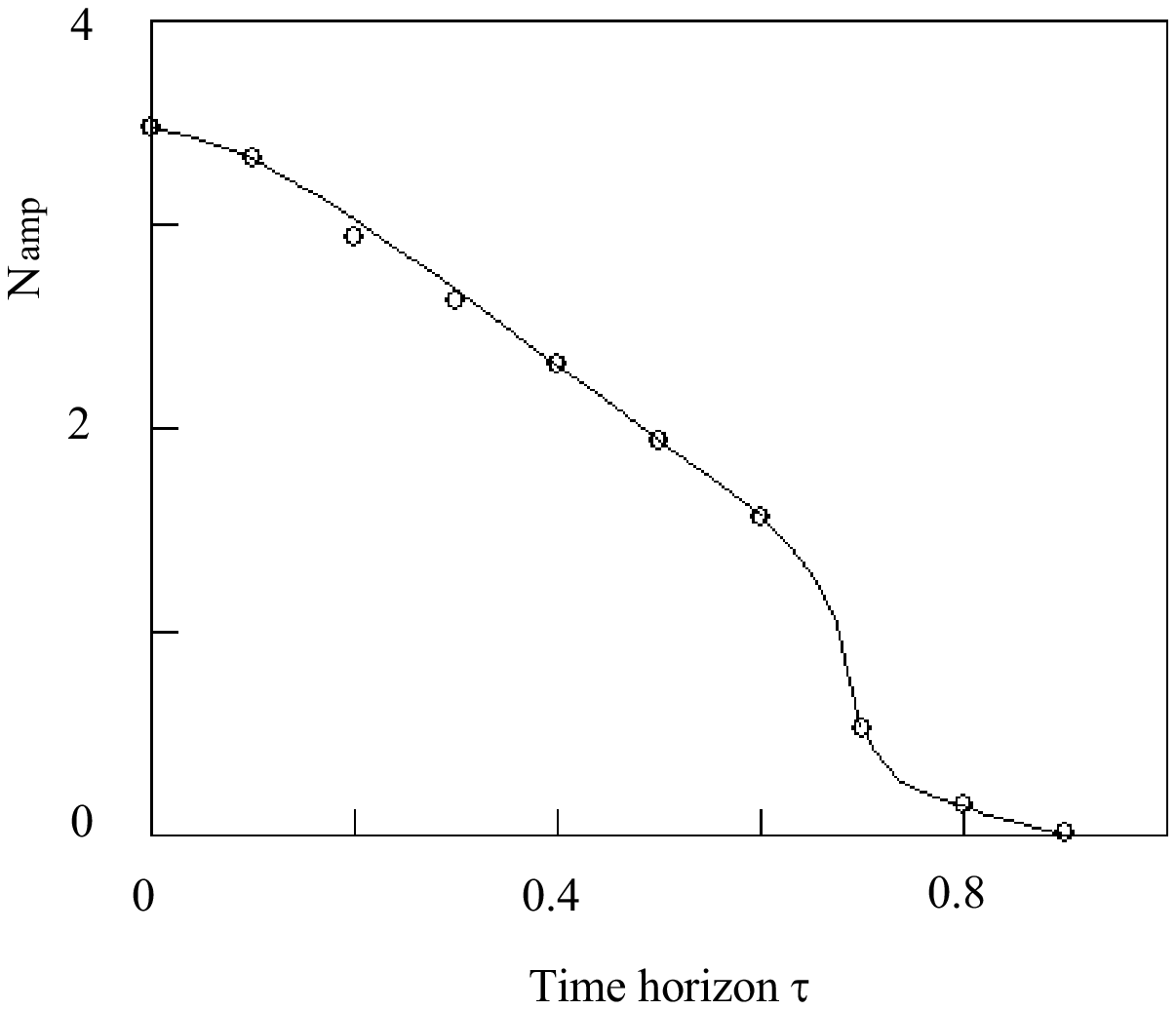}
 \end{center}
\caption[]{ }\label{Fig3}\end{figure}

\begin{figure}[htbp]
 \begin{center}
  \includegraphics[width=10cm]{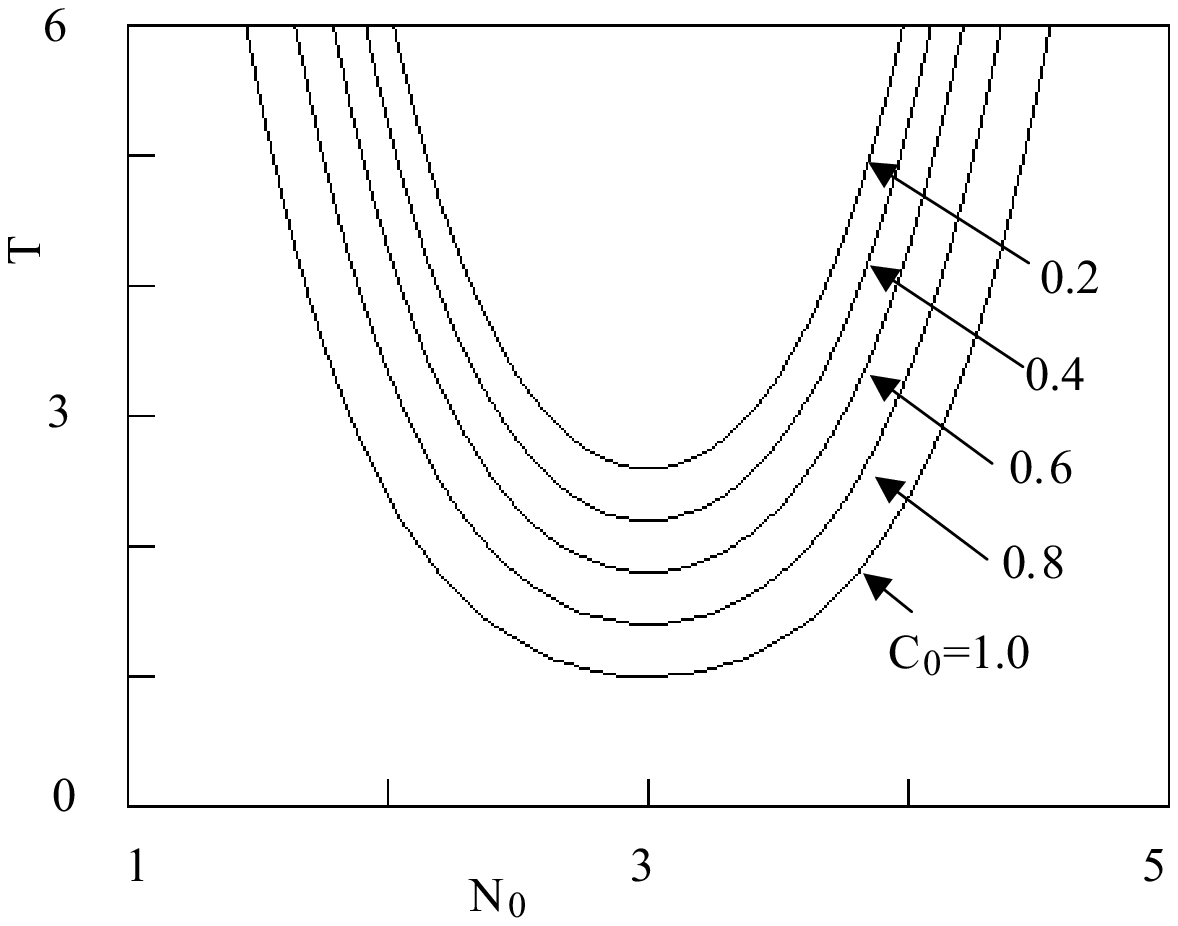}
 \end{center}
\caption[]{ }\label{Fig4}\end{figure}

\begin{figure}[htbp]
 \begin{center}
  \includegraphics[width=10cm]{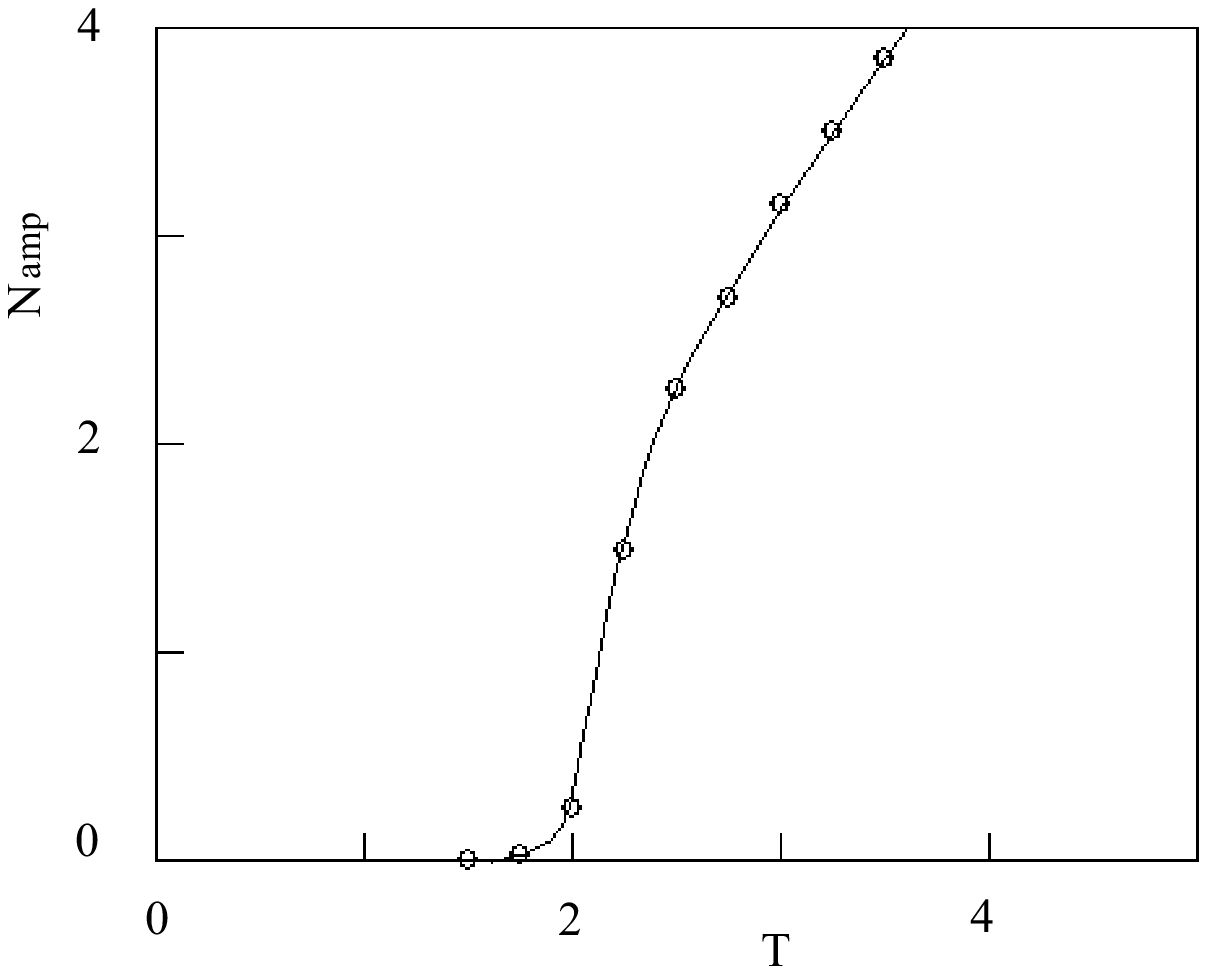}
 \end{center}
\caption[]{ }\label{Fig5}\end{figure}

\begin{figure}[htbp]
 \begin{center}
  \includegraphics[width=10cm]{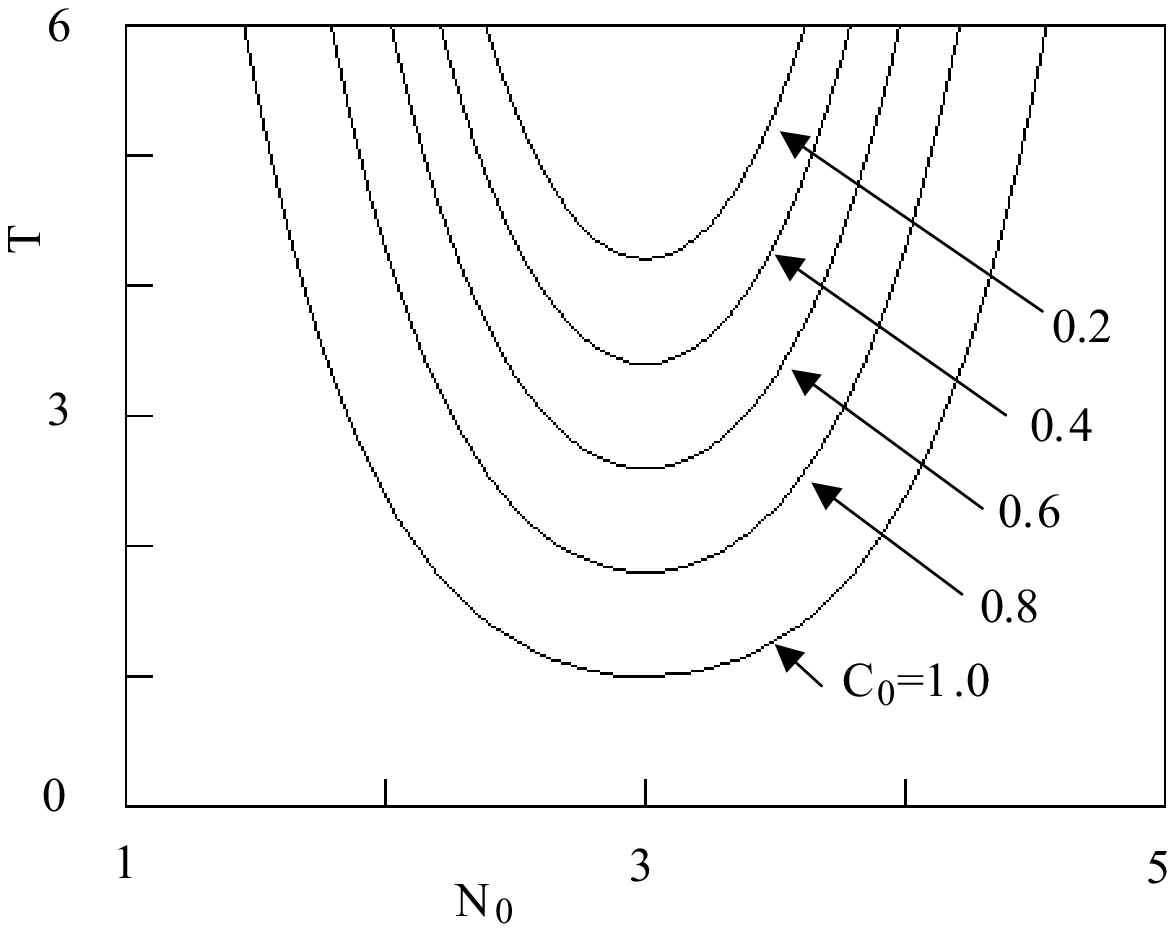}
 \end{center}
\caption[]{ }\label{Fig6}\end{figure}

\begin{figure}[htbp]
 \begin{center}
  \includegraphics[width=10cm]{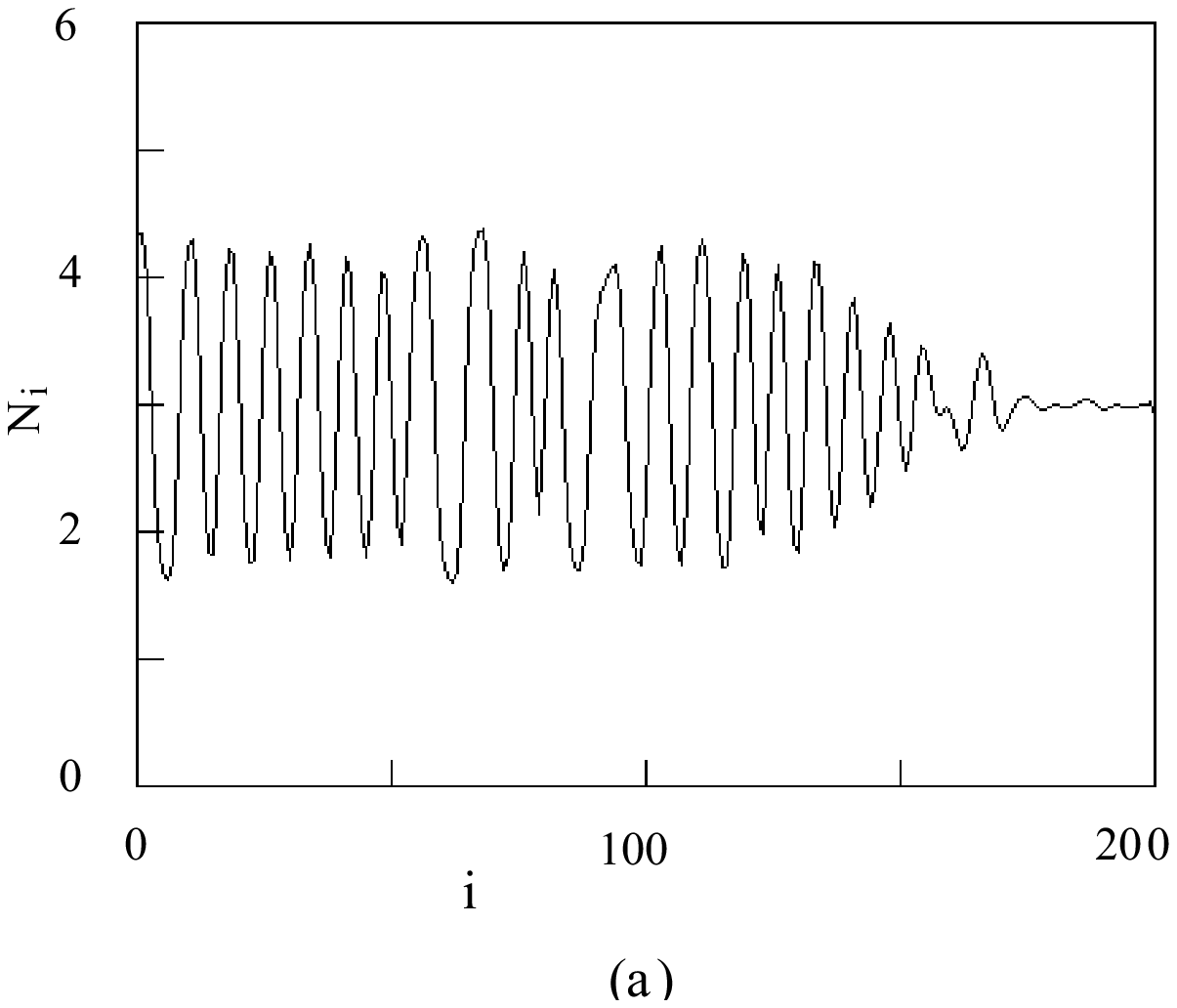}\\[-10cm]
%
  \includegraphics[width=10cm]{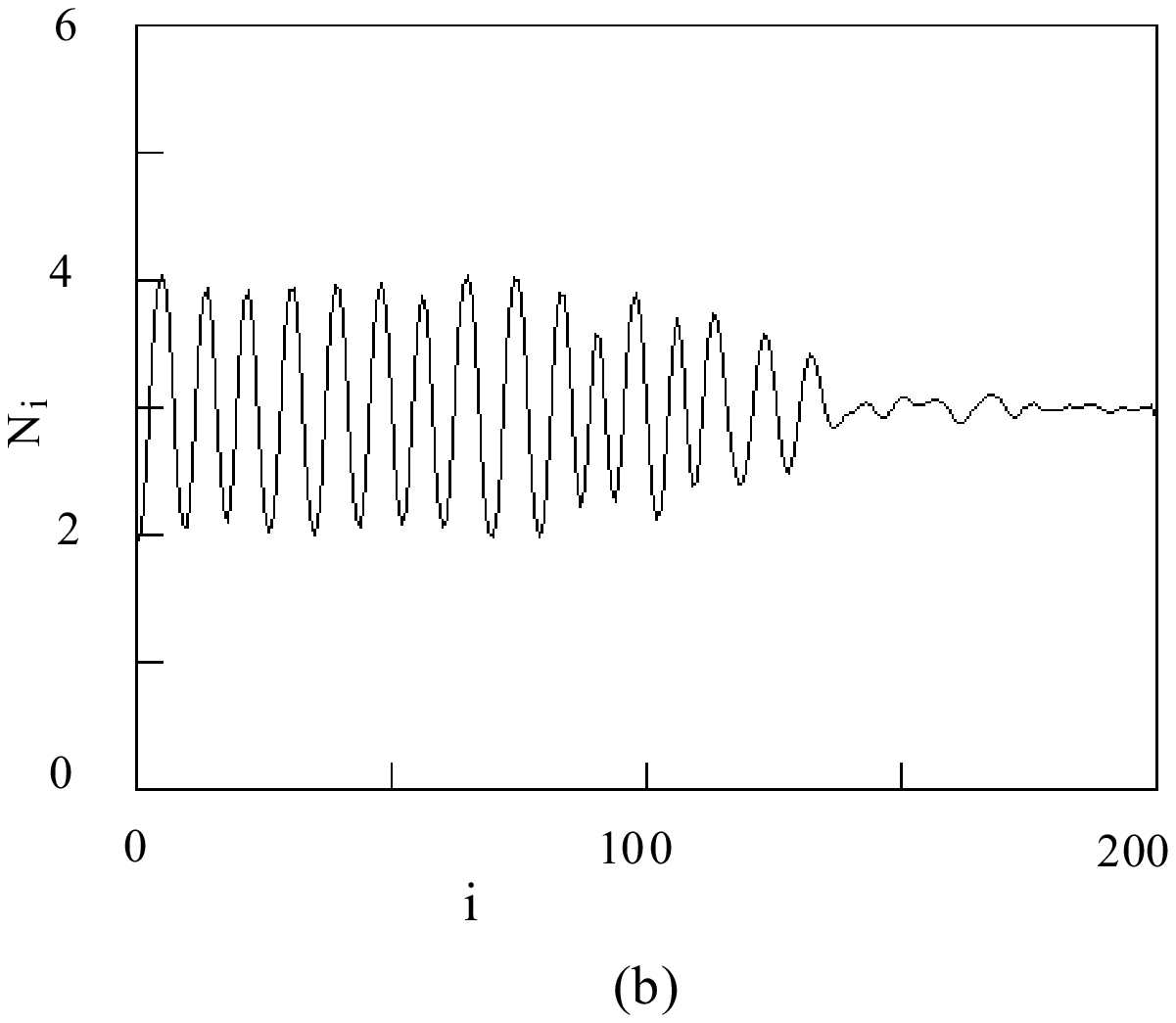}\\[-10cm]
\end{center}
\end{figure}
\newpage
%
\begin{figure}[htbp]
\begin{center}
  \includegraphics[width=10cm]{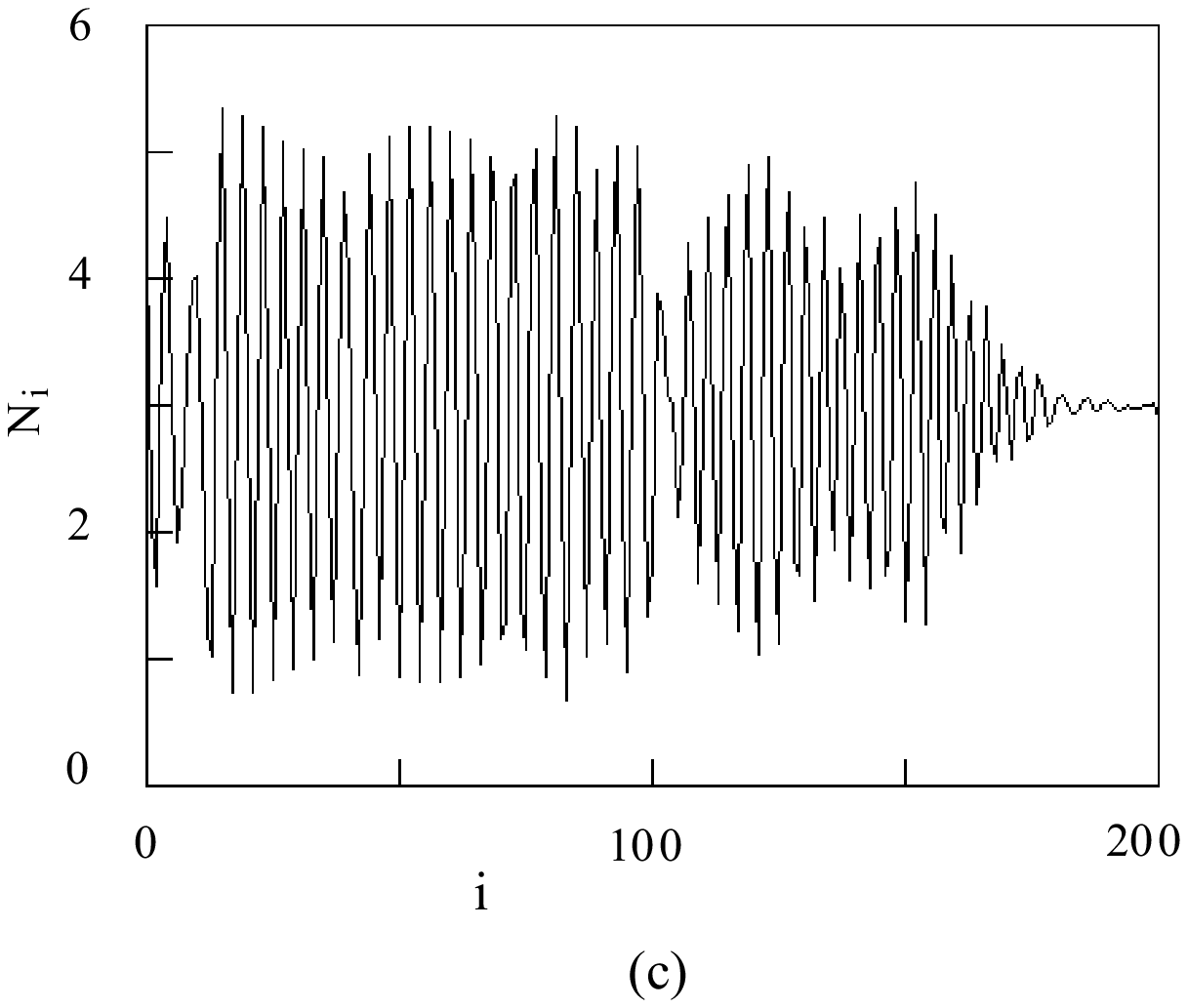}\\[-10cm]
%
  \includegraphics[width=10cm]{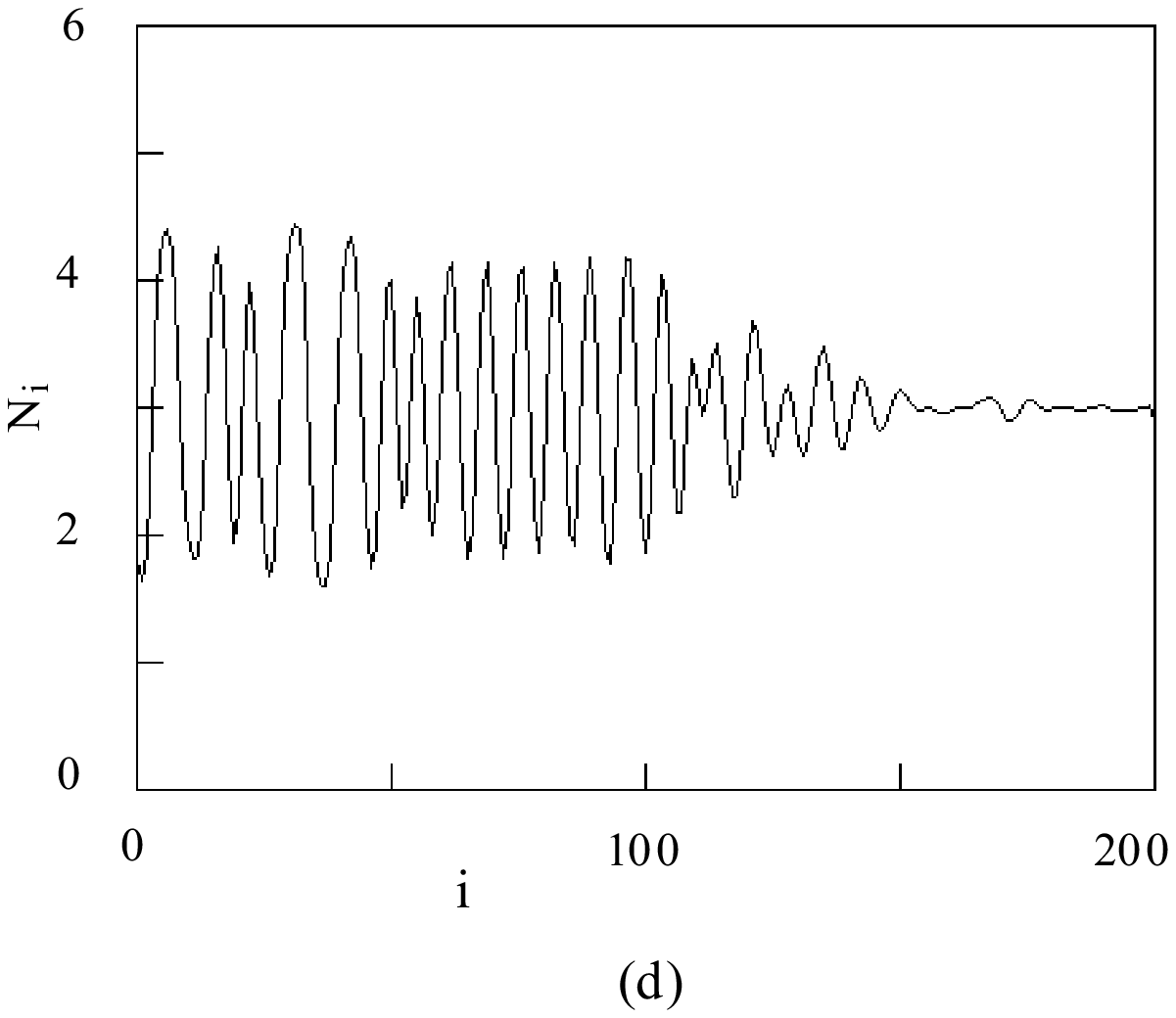}\\[-10cm]
 \end{center}
\caption[]{ }\label{Fig7}\end{figure}

\begin{figure}[htbp]
 \begin{center}
  \includegraphics[width=10cm]{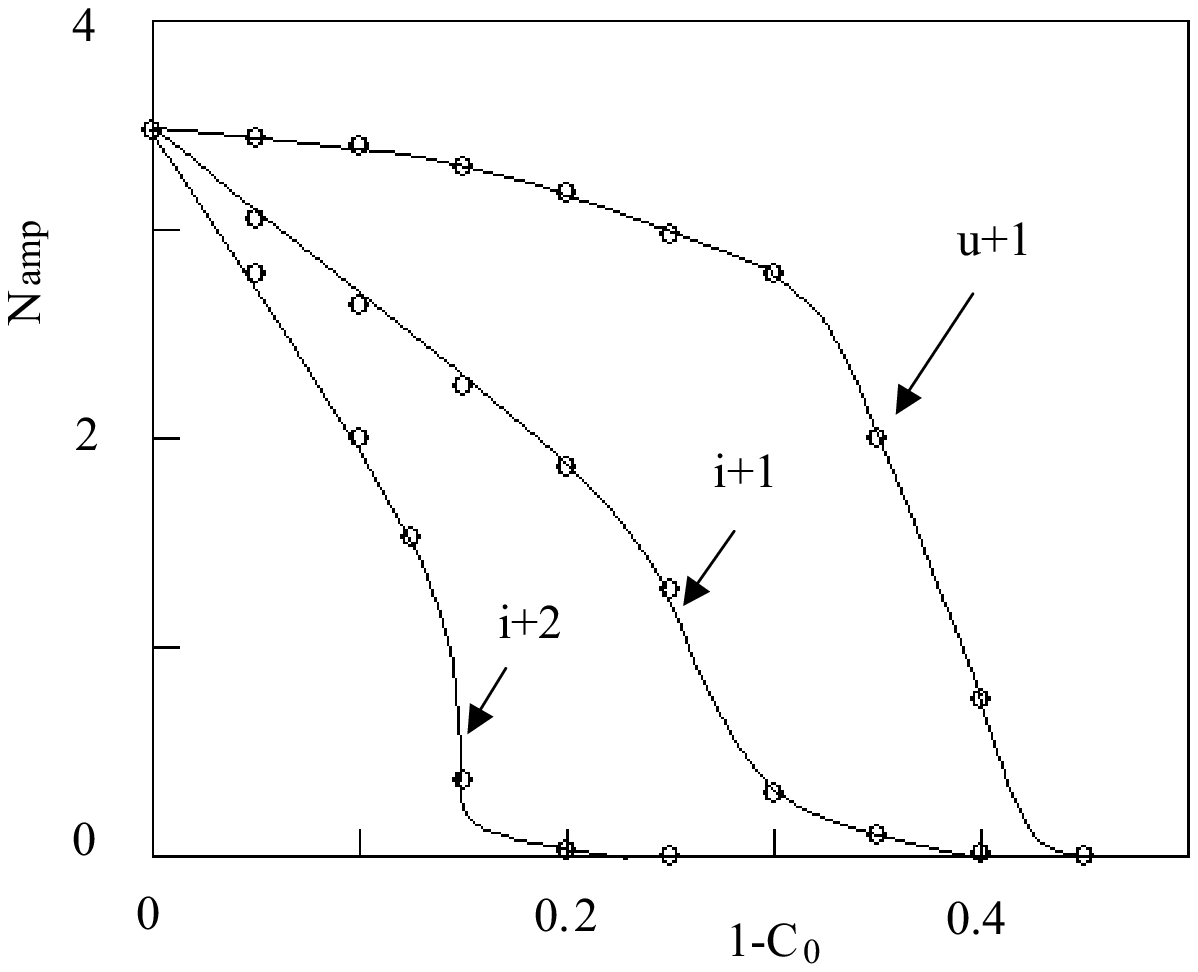}
 \end{center}
\caption[]{ }\label{Fig8}\end{figure}

\end{document}